\documentclass[10pt]{sigplanconf}

\usepackage{amsmath}
\usepackage{fancyvrb}
\usepackage{graphicx}
\usepackage{hyperref}

\newcommand{\logsize}{\footnotesize}
\newcommand{\logwidth}{0.95\columnwidth}

\begin{document}

\conferenceinfo{HotSWUp'08,} {October 20, 2008, Nashville, Tennessee, USA.}
\copyrightyear{2008}
\copyrightdata{ISBN 978-1-60558-304-4/08/10}

\titlebanner{DRAFT---Do not distribute}
\preprintfooter{HotSWUp paper about package upgrades \& rollbacks in FOSS
 distributions}

\title{Package Upgrades in FOSS Distributions: Details and
  Challenges\titlenote{Partially supported by the European Community's
    7th Framework Programme (FP7/2007-2013), grant agreement
    n${}^\circ$214898.}}

\authorinfo{Roberto Di Cosmo \and Stefano Zacchiroli}
 {Universit\'e Paris Diderot, PPS, UMR 7126, France}
 {\{ \href{mailto:roberto@dicosmo.org}{dicosmo} ,
     \href{mailto:zack@pps.jussieu.fr}{zack} \}@pps.jussieu.fr}
\authorinfo{Paulo Trezentos}
 {UNIDE / ISCTE, 1600-082 Lisbon, Portugal}
 {\href{mailto:Paulo.Trezentos@iscte.pt}{Paulo.Trezentos@iscte.pt}}
        
\maketitle

\begin{abstract}
 The \emph{upgrade problems} faced by Free and Open Source Software
 distributions have characteristics not easily found elsewhere. We
 describe the structure of \emph{packages} and their role in the
 \emph{upgrade process}. We show that state of the art package
 managers have shortcomings inhibiting their ability to cope with
 frequent upgrade failures. We survey current countermeasures to such
 failures, argue that they are not satisfactory, and sketch
 alternative solutions.
\end{abstract}


\section{Introduction}
\label{sec:intro}

Free and Open Source Software (\emph{FOSS}) has attracted the
attention of software engineers in the past
decade~\cite{stamelos:foss-code-quality,payne:foss-security} due to
its peculiarities. Among them, release
management~\cite{michlmayr:release-management} is the most relevant
for software upgrades: software bundles---like an operating system
with a basic stack of applications---in the FOSS bazaar are made of
components developed and released independently without a priori
coordination or central authority able to control the involved
parties~\cite{michlmayr:managing-volunteer}. The volunteer nature, the
licensing terms, and the need to reuse, have produced a huge amount of
components, which is unparalleled in the proprietary software
world. Interactions among such components are non trivial, and this is
the main reason why early approaches to software upgrades, where users
had to manually download, compile, install, etc., were doomed to fail.
FOSS \emph{distributions} were therefore introduced around 15 years
ago to reduce the complexity of installations and upgrades for final
users. Distribution maintainers act as intermediaries between
``upstream'' software authors and users, by encapsulating software
components within abstractions called \emph{packages}.

Distributions have been really successful: nowadays every GNU/Linux
user is running one of the hundreds of available distributions. Still,
distributions have inherited some properties from the FOSS bazaar:
complex inter-package dependencies and frequently available package
upgrades. Sysadmins unsurprisingly perform package upgrades at least
once a month~\cite{crameri-mirage}: \emph{new software requirements}
of users require new programs to be installed and old programs to be
removed; \emph{routine upgrades} are required regularly to address
security issues, bug fixes, or to add new features; \emph{release-wide
  upgrades} are less frequent (typically once or twice a year), but
can have higher impact as a significant fraction of the installed
packages are involved. Since a large part of FOSS software is
installed from packages, the damages caused by failed upgrades are
potentially higher than in proprietary systems.

This paper is structured around three claims. The first one is that
\emph{FOSS package upgrades have underestimated peculiarities}. The
claim is supported by Section~\ref{sec:foss-package-upgrade} which
reviews all the actors involved in FOSS package upgrades---packages as
the involved entities, the upgrade process and its decomposition in
clear-cut phases, the possible failures which can occur during
upgrades---and highlights their peculiarities. This paper provides a
detailed description of FOSS package upgrade which, to the best of our
knowledge, was missing from the literature.

The second claim is that \emph{current rollback and snapshot
  techniques are not enough to cope with unpredictable package upgrade
  failures}. Rollback and snapshot techniques are the only
countermeasures currently being proposed against upgrade
failures. Exploiting filesystem-level details, snapshots can be taken
before performing an upgrade, and the possibility to rollback to them
can then be offered to users in case of
failures. Section~\ref{sec:rollback-state} gives an overview of
mainstream snapshot and rollback solutions and argues that no such
solution is satisfactory (in part because of excessive disk-space
requirements induced by long-term upgrade rollbacks, and in part
because it is not always possible to define the rollback scope, i.e.,
what should be rolled-back and what should not).

The third claim is that some of these problems can be tackled by: (1)
adopting lightweight rollback techniques to address short-term
rollback needs, and integrating version control system within package
managers to better handle system-wide configuration files; (2)
designing a domain specific language, equipped with an undo semantics,
for the implementation of scripts which are executed during the
upgrade process and which have side-effects outside the control of
package managers. Details are given in Section~\ref{sec:undo} which
also highlights a novel problem in the specification of upgrade
requests, namely the need of expressing preferences to discriminate
among package statuses which are equivalent from the point of view of
dependency soundness, but which can be sensibly different from that of
user-specified criteria.

\section{Packages, upgrades, failures}
\label{sec:foss-package-upgrade}

\emph{Packages} are abstractions defining the granularity at which
users can act (add, remove, upgrade, etc.) on available software. A
\emph{distribution} is a collection of packages maintained (hopefully)
coherently. The subset of a distribution corresponding to the actual
packages installed on a machine is called \emph{package status} and is
meant to be altered with the \emph{package managers} offered by a
given distribution. They can be classified in two categories:
\emph{installers} which deploy individual packages on the filesystem,
possibly aborting the operation if problems (e.g., unsatisfied
dependencies) are encountered, and \emph{meta-installers} which act at
the inter-package level, solving dependencies and conflicts, and
retrieving packages from remote repositories as needed; \texttt{dpkg}
and \texttt{rpm} are representative examples of installers,
\texttt{apt} and \texttt{urpmi} of meta-installers. We use the term
\emph{upgrade problem} to refer generically to any request to change
the package status; such problems are usually solved by
meta-installers.

\paragraph{Packages} Abstracting over format-specific
details, a \emph{package} is a bundle of 3 main parts:
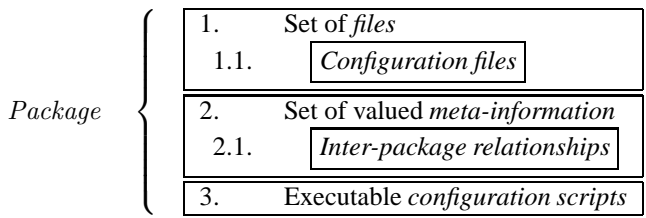
\[
  \mathit{Package} \quad \left\{ \quad
  \begin{tabular}{|ll|}
    \hline
    1. & Set of \emph{files} \\
    ~ 1.1. & ~ ~ \fbox{\emph{Configuration files}} \\[1ex]
    \hline\hline
    2. &  Set of valued \emph{meta-information} \\
    ~ 2.1. & ~ ~ \fbox{\emph{Inter-package relationships}} \\[1ex]
    \hline\hline
    3. &  Executable \emph{configuration scripts} \\
    \hline
  \end{tabular} \right.
  \]
The set of files (1) is common in all software packaging solutions, it
is the filesystem encoding of what the package is delivering:
executable binaries, data, documentation, etc.

\emph{Configuration files} (1.1) is a distinguished subset of shipped
files, identifying those affecting the runtime behavior of the package
and meant to be locally customized with or without package manager
mediation.  Configuration files need to be present in the bundle
(e.g., to provide sane defaults or documentation), but need special
treatment: during installation of new versions of a package, they
cannot be simply overwritten, as they may contain local changes.

Package meta-information (2) contains information which varies from
distribution to distribution. A common core provides: a unique
identifier, software version, maintainer and package description, but
most notably, distributions use meta-information to declare
\emph{inter-package relationships} (2.1). The relationship kinds vary
with the installer, but there exists a de facto common subset
including: dependencies (the need of other packages to work properly),
conflicts (the inability of being co-installed with other packages),
feature provisions (the ability to declare named features as provided
by a given package, so that other packages can depend on them), and
restricted boolean combinations of
them~\cite{edos-package-management}.

Packages come with a set of executable configuration (or
\emph{maintainer}) scripts (3). Their purpose is to let package
maintainers attach actions to hooks executed by the installer; actions
usually come as POSIX shell scripts.

Three aspects of maintainer scripts are noteworthy: (a) they are
ordinary programs that can do anything permitted to the installer
(usually run with administrator rights); (b) the functionality of
maintainer scripts can not be obtained by just shipping extra files:
the scripts may customize part of the package using data which is
available only in the target installation machine, and not necessarily
in the package itself; sometimes the same result obtained using
scripts can be precomputed (increasing package size), sometimes it can
not; (c) maintainer scripts are required to work ``properly'': upgrade
runs in which they fail trigger upgrade failures.

\begin{table}[t]
 \begin{tabular}{l}
  \begin{minipage}[c]{\logwidth}
   \begin{Verbatim}[commandchars=\\\{\},fontsize=\logsize]
# apt-get install aterm\hfill\emph{1. user request}
   \end{Verbatim}
  \end{minipage}
  \\\hline
  \begin{minipage}[c]{\logwidth}
   \begin{Verbatim}[commandchars=\\\{\},fontsize=\logsize]
Reading package lists... Done\hfill\emph{2. dep.resolution}
Building dependency tree... Done
The following extra packages will be installed:
  libafterimage0
0 upgraded, 2 newly installed, 0 to remove and
  1786 not upgraded.
Need to get 386kB of archives.
807kB of additional disk space will be used.
   \end{Verbatim}
  \end{minipage}
  \\\hline
  \begin{minipage}[c]{\logwidth}
   \begin{Verbatim}[commandchars=\\\{\},fontsize=\logsize]
Get: 1 http://ftp.debian.org libafterimage0 2.2.8-2
Get: 2 http://ftp.debian.org aterm 1.0.1-4
Fetched 386kB in 0s (410kB/s)\hfill\emph{3. package retrieval}
   \end{Verbatim}
  \end{minipage}
  \\\hline
  \begin{minipage}[c]{\logwidth}
   \begin{Verbatim}[commandchars=\\\{\},fontsize=\logsize]
    \hfill\emph{5a. (pre-)configuration}
   \end{Verbatim}
  \end{minipage}
  \\\hline
  \begin{minipage}[c]{\logwidth}
   \begin{Verbatim}[commandchars=\\\{\},fontsize=\logsize]
Selecting package libafterimage0.\hfill\emph{4. unpacking}
(Reading database ... 294774 files and dirs installed.)
Unpacking libafterimage0 (libafterimage0_2.2.8-2_i386.deb)
Selecting package aterm.
Unpacking aterm (aterm_1.0.1-4_i386.deb) ...
   \end{Verbatim}
  \end{minipage}
  \\\hline
  \begin{minipage}[c]{\logwidth}
   \begin{Verbatim}[commandchars=\\\{\},fontsize=\logsize]
Setting up libafterimage0 (2.2.8-2) ...
Setting up aterm (1.0.1-4) ...\hfill\emph{5b. (post-)configuration}
   \end{Verbatim}
  \end{minipage}
 \end{tabular}

 \caption{The package upgrade process}
 \label{tab:upgrade-phases}
\end{table}

\paragraph{Upgrades} Table~\ref{tab:upgrade-phases} summarizes the different
phases of what we call the \emph{upgrade process}, using as an example
the popular \texttt{apt} meta-installer (others follow a similar
process).

Phase (1) is a user specification of how she wants the local package
status to be altered. The expressiveness of the language available to
formulate this \emph{user request} varies with the meta-installer: it
can be as simple as requesting the installation/removal of a single
package, or as complex as \texttt{apt} pinning that allows to express
preferences to discriminate among multiple versions of the same
package.

An \emph{upgrade problem} is a triple $\langle U, S_o, R\rangle$,
where $U$ is a distribution (i.e., a set of packages), $S_o\subseteq
U$ is a package status, and $R$ a user request; its \emph{solutions}
are all possible package status $S\subseteq U$,
satisfying:\footnotemark
\begin{enumerate} \renewcommand{\labelenumi}{\alph{enumi}.}
\item The user request $R$ is satisfied by $S$;
\item If $S$ contains a package $p$, it contains all its dependencies;
\item $S$ contains no two conflicting packages;
\item $S$ has been obtained executing all required hooks and none of
  the involved maintainer scripts has failed.
\end{enumerate}
\footnotetext{While (a) is installer-specific, (b) and (c) have been
  generalized and formalized in~\cite{edos2006ase}; studies of (d) are
  still lacking. These are just the \emph{functional} properties of an
  upgrade outcome, but there are also \emph{non-functional} properties
  that can be used to choose \emph{optimal} solutions (e.g.,
  minimality of change, or downtime length); this issue is outside the
  scope of this paper. Note that while checks for (b) and (c) can be
  performed statically, checks for (d) can only be performed at
  run-time while executing scripts.}

Phase (2) performs dependency resolution: it checks whether a package
status satisfying (b) and (c) exists;\footnote{The problem is at least
  NP-complete~\cite{edos-package-management}.} if this is the case one
is chosen in this phase.

Deploying the new status consists of package retrieval (3) and
unpacking (4). Unpacking is the first phase actually changing both the
package status (to keep track of installed packages) and the
filesystem (to add or remove the involved files). During unpacking,
configuration files are treated checking whether local configuration
files have been manually modified or not; if they have, \emph{merging}
is required. The naive solution of asking the user to manually do so
is still the most popular.

Intertwined with package retrieval and unpacking, there are several
configuration phases (5) where maintainer scripts get
executed.\footnote{The details depend on the available hooks;
  \texttt{dpkg} offers: pre/post-unpacking, pre/post-removal, and
  upgrade to some version~\cite{debian-policy}.}

\paragraph{Failures} Each phase of the upgrade process can fail. Dependency
resolution can fail either because the user request is unsatisfiable
(e.g., user error or inconsistent distributions~\cite{edos2006ase}) or
because the meta-installer is unable to find a solution.
Completeness---the guarantee that a solution will be found whenever
one exists---is a desirable meta-installer
property~\cite{mancoosi-debconf8}, unfortunately missing in most
meta-installers, with too few claimed
exceptions~\cite{niemeyer-smart,tucker-opium}.

SAT solving has been proven to be a suitable and complete technique to
solve dependencies~\cite{edos2006ase}, what is still missing is wide
adoption. In that respect recent off-springs\footnote{Apache Maven and
  the Eclipse P2 platform are resorting to SAT solving to manage their
  components and plugins, following the seminal work done by the EDOS
  Project (\url{http://www.edos-project.org}).} are really
promising. \emph{Handling complex user preferences} is a novel problem
for software upgrade. It boils down to letting users specify which
solution to choose among all acceptable solutions. Example of
preferences are policies~\cite{niemeyer-smart,trezentos07}, like
minimizing the download size or prioritizing popular packages, and
also more specific requirements such as blacklisting packages
maintained by an untrusted maintainer.

Package deployment can fail as well. Trivial failures, e.g., network
or disk shortages, can be easily dealt with when considered in
isolation: the whole upgrade process can be aborted and unpack can be
undone, since all the involved files are known; no upgrade is
performed so, the system is unchanged. Maintainer script failures can
not be as easily undone, nor prevented. Scripts are implemented in
Turing-complete languages, and all non-trivial properties about them
are undecidable, including determining \emph{a priori} their effects
to be able to revert them upon failure.

A subtle type of upgrade failure deserves mention: \emph{undetected
  failures}, those failures not observable by the package manager
while the newly installed software can be misbehaving (e.g., a network
service happily restarting after upgrade, but refusing
connections). Undetected failures can take very long (weeks, months)
before being discovered. Often they can be fixed by configuration
tuning, but there are cases in which the desired behavior can no
longer be obtained, leaving upgrade undo as the only solution (in
cases where undoing the upgrade is possible).

\section{Rollback \& snapshot technology overview}
\label{sec:rollback-state}

Current countermeasures to package upgrade failures are based on the
principle of undoing residual effects of failed upgrades. Three
strategies have been proposed: rollbacks, filesystem snapshots, and
purely functional distributions.

\emph{Rollback} capabilities depend on the package manager; the most
well-known implementations are: RPM
transactions~\cite{oden-rpm-transactions} which work at the installer
level, re-creating packages as they are removed, so that they can be
re-installed to undo upgrades; Apt-RPM~\cite{trezentos07} which
implements transactions at the meta-installer level and additionally
handles past versions of configuration files.  All package-based
rollback approaches can track only files which are under package
manager control, and only at package manager invocation time;
therefore none of such approaches can undo maintainer script effects
as they can span the whole system.

\emph{Snapshots} are used to cheaply save copies of physical
filesystems as they were at a given time in the past. ZFS snapshot
(based on \emph{copy on write}) was the first implementation that made
filesystem snapshots popular. ZFS snapshot is integrated with
\texttt{apt-clone} (Nexenta OS meta-installer) to automatically take
snapshots upon upgrades. The Logical Volume Manager (LVM) is a disk
abstraction layer implemented by the Linux kernel, which include
support for copy on write snapshots, without relying on any particular
filesystem implementation.

These snapshot techniques work at the \emph{physical} filesystem
level, hence are unsuitable for recovering from upgrade failures, for
various reasons. The first reason is a granularity mismatch with
package managers that work at the \emph{logical} file system level:
changes induced by upgrades can span several partitions and it can not
be taken for granted that all support snapshots; since even the set of
files of a single package can span multiple partitions, rolling back
only some of them will be too prone to additional problems like
``half-installed'' packages.  How to split the logical filesystem to
support rollbacks is not clear either: while \path{/home} should not
be rolled back (it contains user data), \path{/var} is a hard choice,
since it contains data which are usually affected by maintainer
scripts (and hence needs to be rolled back upon failure) as well as
system logs and database data which usually should not be rolled
back. This problem can be mitigated by a wider acceptance of the
Filesystem Hierarchy
Standard\footnote{\url{http://www.pathname.com/fhs/}} or similar
initiatives to model the purpose of specific paths.

The second reason of the unsuitability of snapshot techniques is disk
usage: even though copy on write requires less space than full
copying, snapshots consume as much space as the divergence between the
snapshot and the live instance. The longer a snapshot is kept alive,
the more physical space is needed to store deltas. Snapshots are then
useful only against quickly discoverable failures (modulo the
filesystem granularity problem), because it cannot be usually afforded
to keep snapshots for the time span of undetected failures.

\emph{Functional distribution} are embodied by
NixOS~\cite{dolstra-nixos} that proposes a functional approach to
package management, where files never change after installation and
are built deterministically evaluating simple functional
expressions. Package deployment is based on garbage collection, hence
packages can never break due to disappearing dependencies. NixOS
suffers from various issues, most notably unconventional configuration
handling intermixed with package building, and the fact that some
actions related to upgrade deployment can not be made purely
functional (e.g., user database management). NixOS made no attempt to
make maintainer scripts purely functional, despite that being the
place where functional purity is needed the most.

\section{Towards perfected package upgrade undo}
\label{sec:undo}

While for detectable failures trade-offs can be made using
snapshot and appropriate partitioning, no fully generic solution
exists to counter upgrade failures. Each of the discussed technologies
focuses on one or more of the axes:
\begin{description}
\item[Domain]: What can and should be undone upon failures (e.g.,
  binary files, configuration files, user files)?
\item[Time]: For how long a specific upgrade can be undone?
\item[Granularity]: Does the undo of one unit imply the undo of other
  units?  Should the unit be file, package, filesystem?
\end{description}
As it is unlikely that a ``one size fits all'' solution exists, we are
pursuing\footnote{In the frame of the Mancoosi project
  (\url{http://www.mancoosi.org})} several research directions to
improve resilience to upgrade failures in FOSS distributions:
\begin{enumerate}
\item Improve meta-installers by the means of (a) lightweight snapshot
  integration and (b) versioning;
\item Define a proper domain specific language (DSL) to be proposed
  as maintainer script implementation language;
\item Define ad-hoc optimized algorithms for handling complex user
  preferences to choose package statuses.
\end{enumerate}
Simple technical improvements can sensibly improve support for upgrade
failures in meta-installers. For example, porting Nexenta ideas to LVM
poses no conceptual problems, and will enable GNU/Linux users to enjoy
similar benefits, no matter the used filesystem.  The need of LVM can
be further relaxed by exploiting lightweight snapshot techniques as
implemented by UnionFS~\cite{wright:unionfs}.

Neither of these two solutions mitigates the problem of long term
snapshot persistence, which will still be too expensive in terms of
disk usage. Hence we also propose to exploit filesystem notifications
(e.g., Linux \texttt{inotify}) to cheaply spot during package upgrades
exactly which files are being modified. This would enable to trim down
snapshots at the end of the upgrade, reducing space requirements.

The need of snapshots can be completely avoided by running upgrades
inside controlled environments as supported by Linux out of the box
(e.g., \texttt{LD\_PRELOAD} to replace the system call library, and
\texttt{ptrace}, a debugging interface to trace process
execution). Using these approaches, one can save on the fly the files
being altered by the upgrade process just before they get
modified~\cite{McQ2005}.

Proper handling of configuration file changes and their undo seems the
easiest goal to achieve, at least at the work-flow level: it is enough
to properly integrate version control systems (VCSs) with
meta-installers.
\texttt{etckeeper}\footnote{\url{http://joey.kitenet.net/code/etckeeper/}}
is a promising example of such an approach. With \texttt{etckeeper} the
whole \path{/etc} directory can be put under version control and enjoy
integration with \texttt{apt} via hooks that commit changes to
configuration files performed by upgrades, so that they can be
recognized and reverted.

This does not address yet the complexity of merging user changes. A
noteworthy example is the need of better integrating the merge
capabilities of modern distributed VCSs. By simply keeping the
pristine configuration files in a separate branch, we can isolate
changes and have a clear view of the differences when manual merge is
required. A related issue is the heterogeneity of languages used to
write configuration files, which inhibits relying on a single
diff/merge tool. To mitigate this problem we observe that for specific
classes of configuration languages (e.g., XML or other structured
syntaxes), syntax-level diff/merge tools can be employed, instead of
the legacy VCS tools, to get rid of bogus merge failures caused by
semantically irrelevant changes.

Regarding maintainer scripts, the only way we see to reliably address
the undo of their effects is by properly formalizing such effects.
Previous attempts to prove properties about shell
scripts~\cite{aiken-sql-injection-detection,mazurak-abash} have given
pale results very far even from the minimal requirement of determining
a priori the set of files touched by their execution, letting aside
how restricted were the considered shell language subsets. Given these
premises, we are skeptical that static analysis can fully solve this
problem. Hence, we are developing a sound DSL equipped with undo
semantics, to be proposed as the implementation language for
maintainer scripts. Although it will be hard to migrate thousands of
existing scripts, empirical analysis on a distribution sample has
shown that most scripts are just a few lines of code, and are mostly
automatically generated. The fact that scripts are maintained by
distribution maintainers will enable us to test-drive the DSL on a
distribution among the Mancoosi partners. The DSL will probably not be
able to address all of maintainer script needs, but if it can handle
most of them, we can resort to other techniques only for the remaining
scripts.

As a first step in DSL design, we are applying fingerprinting
techniques~\cite{aiken:winnowing} to cluster all Debian's maintainer
scripts and get a clear view of all their use cases. It is already
clear that about a half of such scripts only invokes external
idempotent tools to update caches of some data; this class of effects
can be undone by removing the involved data---usually shipped as files
by the owner package---and then re-running the script. What is still
not clear is how heterogeneous are the remaining scripts which escape
the former class.

Considering the intrinsic complexity of the sole dependency
resolution, designing good optimizing algorithms to handle complex
user preferences for package status choices is a rather ambitious
goal. However, the particular shape of inter-package relationships has
enabled deriving rather efficient ad-hoc dependency solvers (e.g.,
\texttt{edos-debcheck}). We believe similar successes can be obtained
for user preferences. Hence we are not only working to apply
multicriteria optimization
techniques~\cite{le-berre:sat-for-dependency}, but also looking for a
tentative ``social'' solution. We are organizing a
competition~\cite{mancoosi-wp1d1} whose participants will compete in
finding the ``best'' algorithm to address the static part of the
upgrade process. We believe the competition has chances to attract
researchers attention, as it will offer real problems collected from
user machines, instead of the usual \emph{in vitro} problems.

\section{Conclusion}
\label{sec:conclusion}

This paper argues that upgrades in FOSS distributions have
underestimated peculiarities. We have discussed the nature of packages
as well as their role in the upgrade process and the potential
failures. We surveyed related work and technologies, showing
their shortcomings, especially in dealing with misbehaving maintainer
scripts. Finally, we presented ongoing research ideas to improve the
state of the art: designing a DSL for implementing maintainer scripts,
and attracting the research community to work on the static part of
package upgrade, including the novel problem of supporting complex
user preferences among packages.

\paragraph{Acknowledgments}
The authors thank the anonymous referees for their feedback; Paulo Trezentos
thanks Ines Lynce and Arlindo Oliveira for interesting discussions on this
topic.

\end{document}